\documentclass[prl,twocolumn,showpacs,amsmath,amssymb,superscriptaddress]{revtex4-1}
\usepackage{graphicx}
\usepackage{epsfig,epstopdf} 
\usepackage{dcolumn}
\usepackage{bm}
\usepackage{bbm}
\usepackage{ulem}
\usepackage{color}
\usepackage{slashed}
\usepackage{hyperref}
\usepackage{mathrsfs}
\usepackage{subfigure}
\usepackage{verbatim}
\usepackage{dsfont}
\usepackage{float}

\newcommand{\nc}{\newcommand}

\nc{\be}{\begin{equation}} \nc{\ee}{\end{equation}}
\nc{\bea}{\begin{eqnarray}} \nc{\eea}{\end{eqnarray}}
\nc{\bean}{\begin{eqnarray*}} \nc{\eean}{\end{eqnarray*}}
\nc{\dg}{\dagger}
\nc{\ua}{\uparrow} \nc{\da}{\downarrow}

\begin{document}

\bibliographystyle{apsrev4-1}

\title{Fingerprints of bosonic symmetry protected topological state in a quantum point contact}
\author{Rui-Xing Zhang and Chao-Xing Liu}
\affiliation{Department of Physics, The Pennsylvania State University, University Park, Pennsylvania 16802}


\begin{abstract}
In this work, we study the transport through a quantum point contact for bosonic helical liquid that exists at the edge of a bilayer graphene under a strong magnetic field. We identify ``smoking gun'' transport signatures to distinguish bosonic symmetry protected topological (BSPT) state from fermionic two-channel quantum spin Hall (QSH) state in this system. In particular, a novel charge insulator/spin conductor phase is found for BSPT state, while either charge insulator/spin insulator or charge conductor/spin conductor phase is expected for the two-channel QSH state. Consequently, a simple transport measurement will reveal the fingerprint of bosonic topological physics in bilayer graphene systems.
\end{abstract}


\maketitle

\textit{Introduction -}
Ever since the discovery of topological insulators (TIs) \cite{fu2007,zhang2009,hasan2010,qi2011}, intensive research has been focused on understanding the role of symmetry in
protecting new topological states, which are known as ``symmetry protected topological (SPT)
states''\cite{chen2012,chen2013}. A grand challenge in this field is to
understand the role of interaction in SPT states and to realize
interacting SPT states in realistic materials.
Recently, it was theoretically proposed that
interaction has a dramatic effect on topological
properties of bilayer graphene under a tilted magnetic field
\cite{bi2016}. The strong magnetic field guarantees the spin conservation, and
drives the system into a quantum spin Hall (QSH) state with edge states described by fermionic two-channel helical Luttinger liquid. Experimentally \cite{young2013},
the two-terminal conductance is found to
approach $\frac{4e^2}{h}$ when chemical potential is tuned into the Zeeman gap between two spin-polarized zeroth Landau levels, which serves as the key signature of helical edge transport in the QSH physics \cite{kane2005a,kane2005b,bernevig2006,konig2007,liu2008,knez2011}.
In Ref. \cite{bi2016}, we analyze the interaction effect in bilayer graphene and demonstrate that fermionic degrees of freedom on the boundary are generally
gapped out. A pair of bosonic edge modes, however, remains gapless
as a result of the symmetry protection of charge conservation ($U(1)_c$ symmetry)
and spin conservation ($U(1)_s$ symmetry). Thus, interactions drive the whole system from a two-channel QSH state into
a bosonic version of topological insulators,
known as bosonic SPT (BSPT) state \cite{chen2012,chen2013,he2016,you2016,yoshida2016}.
Since a pair of dual boson fields of this bosonic edge mode carry
charge-$2e$ excitation and spin-1 excitation, respectively,
and preserve the helical nature, we dub them ``bosonic helical
liquid''. Therefore, bilayer graphene under a strong magnetic field
provides us a unique opportunity to study interacting topological physics
in realistic materials\cite{mazo2014,mazo2015}.

The aim of this work is to explore transport properties of bosonic helical liquid of BSPT state
in bilayer graphene and identify key signatures to distinguish BSPT state from fermionic QSH state.
First of all, the bosonic charge-$2e$ edge excitation of BSPT state carries electric currents and
a two-terminal measurement will also reveal $\frac{4e^2}{h}$ conductance,
taking into account two edges in a realistic sample.
Thus, the two-terminal transport measurements \cite{young2013}
{\it cannot} distinguish the BSPT state from QSH state in bilayer graphene. Several possible experimental probes, such as shot noise measurement of $2e$-charge,
have been considered in Ref. \onlinecite{bi2016}. However, such noise measurement is experimentally challenging
and sometimes controversial, and a simple transport detection of BSPT state is desirable.

\begin{figure}
\centering
\includegraphics[width=0.47\textwidth]{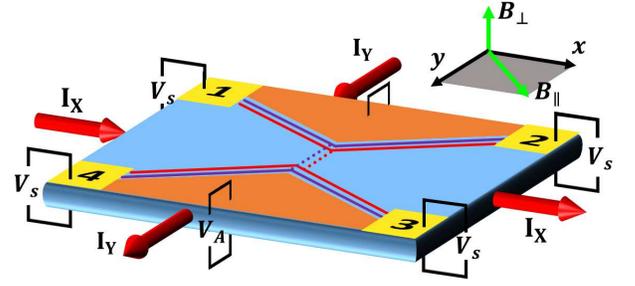}
\caption{QPC setup of a bilayer graphene sample is plotted, where a tilted magnetic field is applied. The BSPT regime is colored in blue, while symmetric potential $V_S$ and asymmetric potential $V_A$ are applied to the yellow and orange regime. Here $V_S$ locally shifts the chemical potential to drive the yellow parts of the sample to be metallic, which thus act as leads.}
\label{Fig:point contact}
\end{figure}

In this work, we study a quantum point contact (QPC) between
two edges of bilayer graphene under a tilted magnetic field, as shown in Fig.
\ref{Fig:point contact}. With the help of this QPC setup,
fingerprints of BSPT state are clearly revealed in the phase diagram of inter-edge tunneling physics.
Based on realistic interaction in bilayer graphene, our main results show (1) a novel charge insulator/spin conductor phase
\cite{hou2009,teo2009}, labelled as IC phase\footnote{Just to clarify, when we talk about BSPT state or fermionic QSH state, we refer to the intrinsic bulk topological state of the system, which is independent of the appearance or absence of QPC structure. When we talk about II/IC/CI/CC phases, we refer to the inter-edge tunneling phase which emerges only when QPC is present.}, when BSPT state is formed, and (2) in contrast, either charge conductor/spin conductor or charge insulator/spin insulator phase, labelled as CC/II phase, for the fermionic two-channel QSH state, where BSPT state is not formed.
Thanks to the unique transport properties in IC phase,
we propose simple two-terminal conductance measurements in both vertical and horizontal directions
in the bilayer graphene QPC.
Perfect insulating behaviors in both directions will be the ``smoking gun'' signal for BSPT physics,
unambiguously distinguishing BSPT state from fermionic QSH state.

\textit{Model Hamiltonian -}
We consider a bilayer graphene sample in a four-terminal configuration
as shown in Fig. \ref{Fig:point contact}.
Both in-plane magnetic field ($B_{\parallel}$)
and out-of-plane magnetic field ($B_{\perp}$) are required to drive the system
into the QSH regime with two-channel helical Luttinger liquid on the boundary \cite{young2013,young2014}.
A strong asymmetric potential ($V_A$) induced by a gate voltage
can drive the system into a layer polarized insulating phase with a trivial gap
\cite{mccann2006,castro2007,kharitonov2012}.
As a result, we can locally gate the sample and
nontrivial edge modes exist at the interface between unbiased region (blue region)
and biased region (orange region),
as shown in Fig. \ref{Fig:point contact}. The local gates can be designed
to form a QPC configuration in this device and the tunneling between two edges
only occurs at the QPC.


As justified in the supplementary materials \cite{supplementary}, helical edge modes can exist in both edges
and are labeled by
the fermionic operators $\psi_{i,l,\lambda}$
that are connected to the lead $i\in\{1,2,3,4\}$ and
characterized by a channel index $l\in\{I,\ II\}$ and a direction index $\lambda\in\{\text{in,\ out}\}$.
Abelian bosonization technique is applied and
the corresponding bosonic chiral fields $\chi_{i,l,\lambda}$ are defined as
$\psi_{i,l,\lambda}=\frac{F_{i,l,\lambda}}{\sqrt{2\pi a_0}}e^{if(\lambda)\sqrt{4\pi}\chi_{i,l,\lambda}}$,
with the Klein factor $F_{i,l,\lambda}$, the coefficient $f(\lambda)=+1 (-1)$ for a right (left) mover and the short-distance cut-off $a_0$. Let us define the edge that connects the leads 1 (3) and 2 (4) as the top (bottom) edge and
the bosonic chiral fields on each edge are related to the $\chi_{i,l,\lambda}$ field by
\bea
\chi_{t(b),l,R}&=&\chi_{1(4),l,out}(-x)\Theta(-x)-\chi_{2(3),l,in}(x)\Theta(x)   \nonumber \\
\chi_{t(b),l,L}&=&\chi_{1(4),l,in}(-x)\Theta(-x)-\chi_{2(3),l,out}(x)\Theta(x),
\label{Eq:Connecting leads}
\eea
with step function $\Theta(x)$. Here the $+x$ direction is defined along the edge from lead 1 (4) to lead 2 (3).
The dual boson fields are introduced as
$\phi_{t(b),l}=\chi_{t(b),l,R}+\chi_{t(b),l,L}$ and
$\theta_{t(b),l}=-\chi_{t(b),l,R}+\chi_{t(b),l,L}$.
Together with the unharmonic terms that respect both $U(1)_c$
and $U(1)_s$ symmetries, the full Hamiltonian is given by
\bea
{\cal H}&=&\sum_{s\in\{t,b\}}\sum_{l=\pm} \frac{v_l}{2}[K_l(\partial_x \phi_{s,l})^2+\frac{1}{K_l}(\partial_x \theta_{s,l})^2] \nonumber \\
&&+g_1\sum_{s}\cos 2\sqrt{2\pi}\phi_{s,-}+g_2\sum_{s}\cos 2\sqrt{2\pi}\theta_{s,-}
\label{Eq:BSPT Hamiltonian}
\eea
where $\phi_{s,\pm}=\frac{1}{\sqrt{2}}(\phi_{s,I}\pm\phi_{s,II})$ and
$\theta_{s,\pm}=\frac{1}{\sqrt{2}}(\theta_{s,I}\pm\theta_{s,II})$
are bonding and anti-bonding fields, respectively.
When $g_1=g_2=0$, this Hamiltonian describes the low-energy edge physics of QSH state with a spin Chern number $2$.
Here $K_{\pm}=\sqrt{\frac{2\pi v_f+2g_5+g_3\pm g_4}{2\pi v_f+2g_5-g_3\mp g_4}}$, and it is expected that $K_->1$. An explicit definition of $g_3$ and $g_4$ can be found in the supplementary materials \cite{supplementary}. A non-zero $g_1$ term is relevant, which
will freeze the $\phi_{s,-}$ field as
$\phi_{s,-}=\frac{(2n_s+1)\pi}{2\sqrt{2\pi}}$ with $n_s\in\mathbb{Z}$, and gap out the anti-bonding boson modes.
The pinning of $\phi_{s,-}$ field is dubbed {\bf BSPT condition},
which mathematically distinguishes bosonic helical liquid from two-channel helical Luttinger liquid.
We further
introduce the notation of spin-charge basis as
\bea
\phi_{\rho}=\phi_{+,+},\ \phi_{\sigma}=\theta_{-,+},\ \theta_{\rho}=\theta_{+,+},\ \theta_{\sigma}=\phi_{-,+}.
\eea
with $\phi_{s=\pm,+}=(\phi_{t,+}\pm\phi_{b,+})/\sqrt{2}$ and
$\theta_{s=\pm,+}=(\theta_{t,+}\pm\theta_{b,+})/\sqrt{2}$.
The corresponding Hamiltonian is
\bea
{\cal H}_\text{BSPT}=\sum_{r=\rho,\sigma}\frac{v_+}{2}[K_+(\partial_x \phi_{r})^2+\frac{1}{K_+}(\partial_x \theta_{r})^2],
\label{Eq:BSPT Hamiltonian-only bonding}
\eea
Therefore, the remaining free bosonic bonding fields $\phi_{s,+}$ and
$\theta_{s,+}$ form helical bosonic edge modes carrying spin-$1$ and charge-$2e$.

\textit{Tunneling physics and Phase diagram -}
For QPC structure, tunneling process is expected to take place
at the contact point $x=0$. Inter-edge tunnelings for a QSH state are only constrained by the symmetries of the system. In a BSPT QPC setup, however, tunneling terms are additionally constrained by the BSPT condition defined above. We will show that this requirement not only constrains the explicit form of tunneling process, but also modifies the scaling dimension of tunneling operators and greatly changes the phase diagram of tunneling process.

Let us start with the single-particle tunneling, and $U(1)_s$ symmetry requires that an electron must switch its velocity when hopping between different edges. Generally, the single-particle tunneling operator 
is
\bea
T_{l,l'}&=&t_{l,l'}\psi^{\dagger}_{t,l,L}\psi_{b,l',R}+h.c.
\label{Eq:Single particle tunneling}
\eea
In the bosonized language, $T_{l,l'}=t_{l,l'}\cos \sqrt{\pi}[\phi_{+,+}+\theta_{-,+}-f_+(\phi_{+,-}+\theta_{-,-})-f_-(\phi_{-,-}+\theta_{+,-})]$, where $f_{\pm}=\frac{1}{2}[(-1)^l\pm (-1)^{l'}]$. BSPT condition guarantees that
the correlation function of its dual fields $\langle\theta_{s,-}(\tau)\theta_{s,-}(0)\rangle$ diverges as $g_1\rightarrow\infty$ \cite{supplementary}. As a result, the correlation function of any vertex operator of $\theta_{s,-}$ vanishes since
$\langle e^{i\alpha\theta_{s,-}(\tau)}e^{-i\alpha\theta_{s,-}(0))}\rangle=e^{-\frac{\alpha^2}{2}\langle[(\theta_{s,-}(\tau)-\theta_{s,-}(0))^2]\rangle}$.
This immediately implies that any vertex operator of $\theta_{s,-}$ is vanishing under RG operation.
Since $\theta_{s,-}$ always appears in $T_{l,l'}$,
we conclude that single particle tunneling $T_{l,l'}$ is generally forbidden in the BSPT QPC.
Physically, this implies that single-particle tunneling is incompatible with the BSPT condition, and violates the
bosonic nature of BSPT state.

\begin{figure}
\centering
\includegraphics[width=0.5\textwidth]{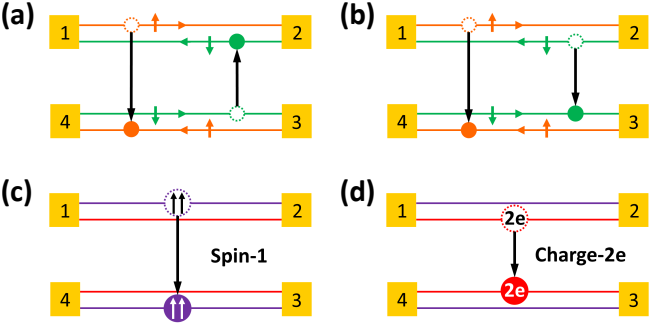}
\caption{Two-particle tunneling processes of spin-1 tunneling and charge-$2e$ tunneling are plotted: (i) (a) and (b) in the fermion limit; (ii) (c) and (d) in the BSPT limit.}
\label{Fig:tunneling}
\end{figure}

Next, we examine the two-particle tunneling shown in Fig. \ref{Fig:tunneling} (a), where a right mover on the top edge (spin-up) tunnels to a left mover on the bottom edge (spin-up), and a right mover on the bottom edge (spin-down) simultaneously tunnels to a left mover on the top edge (spin-down). As a result, the charge transfer between the top and bottom edges is zero, while the spin transfer is one. This type of spin-1 tunneling process is mathematically described by
\bea
V^{\sigma}=v^{\sigma}_{l_1,l_2,l_3,l_4}\psi^{\dagger}_{b,L,l_1}\psi_{t,R,l_2}\psi^{\dagger}_{t,L,l_3}\psi_{b,R,l_4}+h.c.,
\label{Eq:Spin-1 tunneling}
\eea
where $l_{1,2,3,4}\in{I,II}$. Under BSPT condition, the absence of anti-bonding field $\theta_{s,-}$ in $V^{\sigma}$ yields a strong constraint on the channel index $l_i$: $l_1=l_4=l,\ \ l_2=l_3=l'$, which leads to
\bea
V^{\sigma}=v^{\sigma}\cos2\sqrt{\pi}\phi_{+,+}=v^{\sigma}\cos2\sqrt{\pi}\phi_{\rho}.
\label{Eq:Spin-1 general}
\eea
There exists another type of symmetry allowed two-particle tunneling term, which describes inter-edge transfer of $2e$ charge and zero spin, as shown in Fig. \ref{Fig:tunneling} (b):
\bea
V^{\rho}=v^{\rho}_{l_1,l_2,l_3,l_4}\psi^{\dagger}_{b,L,l_1}\psi_{t,R,l_2}\psi^{\dagger}_{b,R,l_4}\psi_{t,L,l_3}+h.c.
\eea
The condition for a non-vanishing $V^{\rho}$ can be similarly identified as $l_1\neq l_4,\ \ l_2\neq l_3$, leading to the following bosonized expression of charge-$2e$ tunneling as
\bea
V^{\rho}=v^{\rho}\cos 2\sqrt{\pi}\theta_{-,+}=v^{\rho}\cos 2\sqrt{\pi}\phi_{\sigma}.
\label{Eq:Charge-2e general}
\eea

As shown in Ref. \cite{bi2016}, the elementary bosonic excitations on the edge $s$ are found to be either charge-$2e$ spin-singlet Cooper pair $\Phi_{s,q=2e}=\psi_{s,I,R}\psi_{s,II,L}-\psi_{s,I,L}\psi_{s,II,R}\sim e^{-i\sqrt{2\pi}\theta_{s,+}}$ or spin-1 chargeless spinon $\Phi_{s,\sigma=1}=\psi^{\dagger}_{s,I,\da}\psi_{s,I,\ua}-\psi^{\dagger}_{s,II,\da}\psi_{s,II,\ua}\sim e^{-i(-1)^s\sqrt{2\pi}\phi_{s,+}}$. For the definition of bosonic operator $\Phi_{s,\sigma=1}$, we have used the convention $(-1)^t=-1$ and $(-1)^b=1$, which originates from opposite spin-momentum locking at different edges.
The above two-particle tunneling terms
can be rewritten as,
\bea
V^{\sigma}&=&v^{\sigma}\Phi_{b,\sigma=1}^{\dagger}\Phi_{t,\sigma=1}+h.c. \nonumber \\
V^{\rho}&=&v^{\rho}\Phi_{b,q=2e}^{\dagger}\Phi_{t,q=2e}+h.c.
\label{Eq:AFM and Cooper pair tunneling}
\eea
Therefore, two-particle tunneling $V_{\sigma}$ and $V_{\rho}$ are physically interpreted as the tunneling of bosonic quasi-particles across the QPC, as shown in Fig. \ref{Fig:tunneling} (c) and (d). In other words, Eq. (\ref{Eq:AFM and Cooper pair tunneling}) demonstrates the minimal tunneling events allowed in a {\it bosonic} SPT system.


Now we are ready to analyze and compare the phase diagram of tunneling physics
for bilayer graphene QPC structure with and without the formation of BSPT state. In a series of pioneering works, the QPC physics of fermionic 1-channel helical Luttinger liquid and fermionic 4-channel helical Luttinger liquid have been studied in a QSH system \cite{hou2009,strom2009,teo2009} and a bilayer graphene with domain walls \cite{wieder2015}. The phase diagram of our bilayer graphene QSH state follows the paradigm in the above systems: (1) In the weak interaction limit, both single-particle and two-particle tunneling terms are small and irrelevant, which defines CC phase. However, a duality transformation of CC phase reveals another stable fixed point where the QPC is pinched off, giving rise to the so-called II phase \cite{teo2009}. Therefore, CC and II fixed points are separated by a QPC pinch-off transition in this parameter regime. (2) As the repulsive (attractive) interaction strengths exceed critical values, QPC is driven into the IC (CI or charge conductor/spin insulator) phase where spin-1 (charge-$2e$) tunneling is relevant. We have mapped out
the phase diagram of fermionic two-channel QSH state in QPC setup of bilayer graphene, as shown in Fig. \ref{Fig:Phase diagram} (a). More details can be found in the supplementary materials \cite{supplementary}.


When bulk BSPT state is formed, however, BSPT condition freezes the anti-bonding degree of freedom and removes the role of $K_-$ in the phase diagram. Scaling dimensions of two-particle tunneling terms are further modified to $\Delta(v^{\sigma})=\frac{1}{K_+}$ and $\Delta(v^{\sigma})=K_+$, in comparison to the QSH case \cite{supplementary}.
This change of scaling dimensions leads to different RG equations
\bea
\frac{dv^{\sigma}}{da}=(1-\frac{1}{K_+})v^{\sigma},\ \ \frac{dv^{\rho}}{da}=(1-K_+)v^{\rho},
\label{Eq:RG}
\eea
with real space scaling factor $a$ for $v^{\sigma,\rho}$.
For $K_+>1$, we find $v^{\sigma}$ is relevant while
$v^{\rho}$ is irrelevant, leading to the IC phase. In contrast, the CI phase appears for
$K_+<1$ and is separated from the IC phase by a critical point at $K_+=1$,
as shown in Fig. \ref{Fig:Phase diagram} (b).
Comparing Fig. \ref{Fig:Phase diagram} (a) and (b), we find two phase diagrams
are completely different in the weak interaction limit $K_+\approx 1$, thus
providing a route to distinguish BSPT state and fermionic two-channel
QSH state in bilayer graphene.

\begin{figure}[t]
\centering
\includegraphics[width=0.47\textwidth]{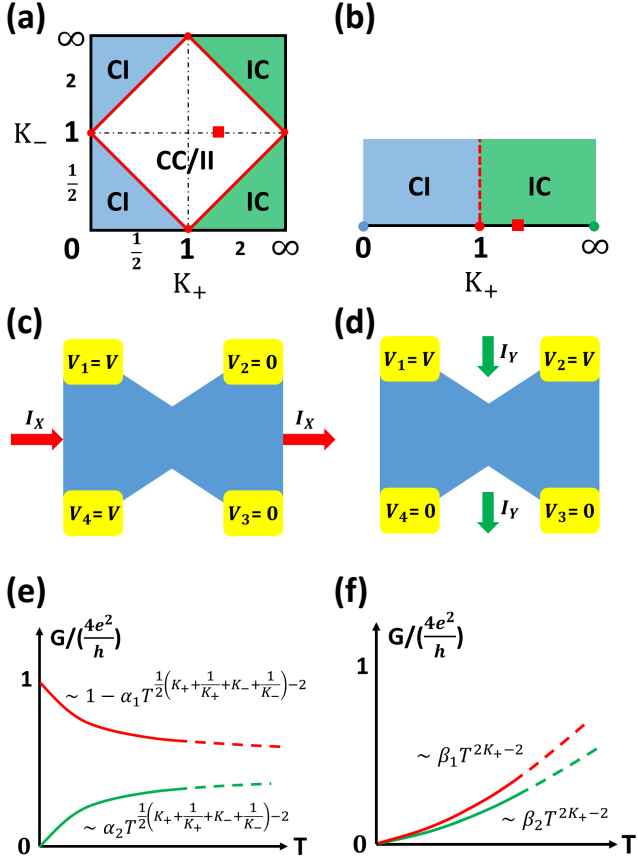}
\caption{Phase diagram of the QPC physics is plotted for: (i) two-channel QSH state in (a); (ii)  BSPT state in (b). Voltage configurations of the proposed two-terminal measurement are shown in (c) and (d). Temperature dependence of $G_{XX}$ (red line) and $G_{YY}$ (green line) are also plotted for QSH state in (e) and BSPT state in (f).}
\label{Fig:Phase diagram}
\end{figure}

\textit{Experimental detection -}
Based on the phase diagram (Fig. \ref{Fig:Phase diagram} (a) and (b)), we next turn to realistic bilayer graphene systems.
First, we need to give an estimate of the Luttinger parameters $K_{\pm}$, which can be extracted from
the screened Coulomb interaction between two edge state electrons.
As discussed in the supplementary materials \cite{supplementary}, after mapping the screened Coulomb interaction into the four-fermion interactions
in Luttinger liquids, we find that $K_+$ is determined by the ratio between interaction strength and
kinetic energy of the edge modes, while $K_-$ is related to the difference between intra- and inter-Landau level
interactions. Assuming the out-of-plane magnetic field to be $2$ Tesla and a substrate dielectric constant $\epsilon=5$,
we find that $K_+=1.43$ and $K_-=1.02$ \footnote{In our work, Luttinger parameter $K_{\pm}$ is defined  in analogous to the inverse of Luttinger parameter $g$ in Ref. \cite{teo2009}. To be specific, repulsive interaction implies $K_+>1$ in our notation and $g<1$ in Ref. \cite{teo2009}.} in our bilayer graphene system, which is depicted by the red square
in both Fig. \ref{Fig:Phase diagram} (a) and (b). Based on this estimate, we conclude that the formation of BSPT state
drives the QPC system in bilayer grahene from CC/II phase into IC phase. In other words,
probing IC phase in the QPC can serve as the transport evidence of the BSPT state in bilayer graphene.

In the following, we demonstrate that a simple transport measurement will unambiguously
distinguish IC phase from CC/II phase. We consider to apply either horizontal ($V_X=V_1-V_2-V_3+V_4$)
or vertical bias voltages ($V_Y=V_1+V_2-V_3-V_4$).
The simplest voltage configurations are shown in Fig. \ref{Fig:Phase diagram} (c) and (d),
which are effectively two-terminal setups in two orthogonal directions.
A horizontal current $I_X=I_1-I_2-I_3+I_4$ and a vertical current $I_Y=I_1+I_2-I_3-I_4$
can be measured to extract conductances along both directions,
where $I_i$ ($V_i$) is the lead current (voltage)
for lead $i\in\{1,2,3,4\}$. The current operators above are related to
the boson current operators as $I_X=I_{\rho}$ and $I_Y+I_{\sigma}=\frac{4e^2}{h}V_Y$.
These relations can be easily verified with the help of Eq. (\ref{Eq:Connecting leads}),
together with the definition of spin/charge current $I_{\rho/\sigma}=-\frac{2}{\sqrt{\pi}}\partial_{t}\phi_{\rho/\sigma}$.
For the CC phase of QSH state, both single-particle tunneling and two-particle tunneling terms are irrelevant,
so $\phi_{\rho}$ and $\phi_{\sigma}$ are free boson fields whose currents are accompanied by a quantized conductance.
This gives rise to $I_{X}=\frac{4e^2}{h}V_{X}$ while $I_{Y}=0$.
From the duality relation between CC and II phases, we immediately find that $I_{Y}=\frac{4e^2}{h}V_{Y}$
and $I_{X}=0$ for II phase. Therefore, a QSH sample is always found to be a perfect conductor
along \textit{either horizontal or vertical direction}, while a perfect insulator
along the corresponding orthogonal direction. On the other hand, for a BSPT system,
the IC phase exhibits relevant spin-1 tunneling process $V^{\sigma}$, which gaps out only $\phi_{\rho}$ field. As a consequence, both $I_X$ and $I_Y$ are vanishing and the current flows in the leads are constrained by $I_1=-I_2=I_3=-I_4$ \cite{hou2009,teo2009}. Thus, the BSPT QPC setup shows the perfect insulating behaviors in \textit{both horizontal and vertical directions}! This simple and feasible transport measurement will be the smoking gun evidence of BSPT state.

The distinction between QSH state and BSPT state is further demonstrated when temperature effects are incorporated. Temperature dependence of horizontal conductance $G_{XX}$ (red line) and vertical conductance $G_{YY}$ (green line) are plotted in both CC phase of QSH state (assuming CC phase for QSH state) and IC phase of BSPT state. In the CC phase of QSH state, $G_{XX}$ ($G_{YY}$) experiences a power-law decay (increase) from the plateau value (zero), and the power-law scaling relation reflects the scaling dimension of single-particle tunneling operators \cite{supplementary}. In the IC phase of BSPT state, however, both conductances share a similar power-law increase from zero. In contrast to CC phase, the power of temperature dependence is determined by two-particle (bosonic-particle) tunneling, which only depends on $K_+$. With our previous estimation of $K_+$ and $K_-$, we find $\Delta G_{XX/YY}\sim T^{0.07}$ for QSH state while $\Delta G_{XX/YY}\sim T^{0.86}$ for BSPT state.
Therefore, the temperature scaling of $G_{XX}$ and $G_{YY}$ reflects the tunneling mechanism in the QPC for either QSH state
or BSPT state.




\textit{Conclusion -}
We proposed that a simple QPC setup ``magically'' implements two-terminal transport measurements to unambiguously distinguish BSPT state from QSH state. In particular, QPC reveals the fingerprints of bosonic physics in the phase diagram of inter-edge tunneling physics, and binds BSPT state with exotic IC physics in bilayer graphene systems. We notice that the IC phase
has not been experimentally realized, probably because it requires a strong interaction in conventional QSH systems.
In contrast, our estimate shows that it can be driven by realistic Coulomb interaction in bilayer graphene.
Another great advantage of bilayer graphene is that its QPC can be feasibly designed and controlled by gate voltages, as shown in Fig. \ref{Fig:point contact}, which is absent in other QSH systems.
In the supplementary materials \cite{supplementary}, a detailed calculation of extracting effective charge from shot noise spectrum is also presented. Bosonic $2e$-charge is found, which originates from the instanton tunneling events of IC fixed point. Compared with this direct probe of bosonic electric charge, the transport measurements we proposed are much simpler and more feasible for experiment realization.


\textit{Acknowledgement}
We would like to thank Cenke Xu for useful discussions.
C.-X.L. acknowledge the support from Office of Naval Research
(Grant No. N00014-15-1-2675).

\bibliography{BSPT}

\onecolumngrid
\newpage

\subsection{\large Supplementary Materials for ``Fingerprints of bosonic symmetry protected topological state in a quantum point contact"}

\section{Landau level zero mode in bilayer graphene and helical edge states}
At valley K, the effective Hamiltonian of bilayer graphene under an out-of-plane magnetic field is given by
\bea
H_{K}=
\begin{pmatrix}
-U(x) & v_f\pi_- & 0 & 0 \\
v_f\pi_+ & -U(x) & t & 0 \\
0 & t & U(x) & v_f\pi_- \\
0 & 0 & v_f\pi_+ & U(x) \\
\end{pmatrix}
\eea
Here the bases are $(|A\rangle,|B\rangle,|A'\rangle,|B'\rangle)^T$. We have defined conjugate momentum $\pi_{\pm}=\pi_x\pm i\pi_y$, where $\pi_{x,y}=-i\hbar\partial_{x,y}+eA_{x,y}$. $t$ is the nearest neighbor interlayer hopping between $|B\rangle$ and $|A'\rangle$. The boundary between layer-polarized trivial insulating state and BSPT state is introduced by a electric asymmetric potential $U(x)=U_0\Theta(x)$ which is non-uniform in the $x$ direction, as demonstrated in Fig. 1 in the main text. Experimentally, the spatial dependence of $U(x)$ is actually smooth in the strong magnetic field limit, as its length scale is much larger than the magnetic length $l_B$ which is around $10$ nm. Define the following creation and annihilation operators
\bea
a=\frac{l_B}{\sqrt{2}\hbar}\pi_-,\ a^{\dagger}=\frac{l_B}{\sqrt{2}\hbar}\pi_+,
\eea
with $[a,a^{\dagger}]=1$ and magnetic length $l_B=\sqrt{\frac{\hbar}{eB_z}}\approx\frac{26 \text{nm}}{\sqrt{B}}$. The effective Hamiltonian now becomes
\bea
H_{K}=
\begin{pmatrix}
-U(x) & \frac{\sqrt{2}\hbar v_f}{l_B}a & 0 & 0 \\
\frac{\sqrt{2}\hbar v_f}{l_B}a^{\dagger} & -U(x) & t & 0 \\
0 & t & U(x) & \frac{\sqrt{2}\hbar v_f}{l_B}a \\
0 & 0 & \frac{\sqrt{2}\hbar v_f}{l_B}a^{\dagger} & U(x) \\
\end{pmatrix}
\eea
Introduce $a|\varphi_n\rangle=\sqrt{n}|\varphi_{n-1}\rangle$ and $a^{\dagger}|\varphi_n\rangle=\sqrt{n+1}|\varphi_{n+1}\rangle$, where $|\varphi_n\rangle$ is the $n$th Landau level eigenstate and its explicit form depends on the gauge we choose. The form of $H_K$ inspires us to guess a trial wavefunction $\Psi_n=(c_{1,n}|\varphi_{n-2}\rangle,c_{2,n}|\varphi_{n-1}\rangle,c_{3,n}|\varphi_{n-1}\rangle,c_{4,n}|\varphi_{n}\rangle)^T$. Then the corresponding eigen-state equation is
\bea
\begin{pmatrix}
-U(x) & \frac{\sqrt{2(n-1)}\hbar v_f}{l_B} & 0 & 0 \\
\frac{\sqrt{2(n-1)}\hbar v_f}{l_B} & -U(x) & t & 0 \\
0 & t & U(x) & \frac{\sqrt{2n}\hbar v_f}{l_B} \\
0 & 0 & \frac{\sqrt{2n}\hbar v_f}{l_B} & U(x) \\
\end{pmatrix}
\begin{pmatrix}
c_{1,n} \\ c_{2,n} \\ c_{3,n} \\ c_{4,n}
\end{pmatrix}
=E \begin{pmatrix}
c_{1,n} \\ c_{2,n} \\ c_{3,n} \\ c_{4,n}
\end{pmatrix}.
\eea
For $n=0$ and $n=1$, the ansatz wavefunctions reduce to $\Psi_{0,K}=(0,0,0,|\varphi_{0}\rangle)^T$ and $\Psi_{1,K}=(0,c_{2,1}|\varphi_{0}\rangle,c_{3,1}|\varphi_{0}\rangle,c_{4,1}|\varphi_{1}\rangle)^T$. By solving the eigen-equation exactly in the $U=0$ limit, we find two zero modes with $E=0$ and characterized by
\bea
\Psi_{0,K}&=&(0,0,0,|\varphi_{0}\rangle)^T \nonumber \\
\Psi_{1,K}&=&(0,\frac{1}{\sqrt{1+\gamma^2}}|\varphi_{0}\rangle,0,-\frac{\gamma}{\sqrt{1+\gamma^2}}|\varphi_{1}\rangle)^T,
\eea
where $\gamma=\frac{t}{\sqrt{2}\hbar v_f/l_B}$. For the $K'$ valley, a similar analysis is quite straightforward and we find two more zero modes in the first two Landau levels,
\bea
\Psi_{0,K'}&=&(|\varphi_{0}\rangle,0,0,0)^T \nonumber \\
\Psi_{1,K'}&=&(\frac{\gamma}{\sqrt{1+\gamma^2}}|\varphi_{1}\rangle,0,\frac{1}{\sqrt{1+\gamma^2}}|\varphi_{0}\rangle,0)^T.
\eea
In the weak potential limit, we treat $U(x)$ as a perturbation and find the following low-energy dispersion of the above zero modes:
\bea
E_{0,K}&=&U,\ \ E_{1,K}=U\frac{\gamma^2-1}{\gamma^2+1} \nonumber \\
E_{0,K'}&=&-U,\ E_{1,K'}=-U\frac{\gamma^2-1}{\gamma^2+1}.
\eea
With the realistic parameters for bilayer graphene \cite{mccann2013}, we find that $\gamma\approx \frac{11.37}{\sqrt{B}}=8$ when out-of-plane field $B=2$ Tesla. This gives rise to $\frac{\gamma}{\sqrt{1+\gamma^2}}\approx 0.99$. Therefore, $\Psi_{0,K}$ and $\Psi_{1,K}$ are identified as two right moving edge modes, while $\Psi_{0,K}$ and $\Psi_{1,K}$ are two left moving edge modes. \\

\begin{figure}
  \centering
  \includegraphics[width=\textwidth]{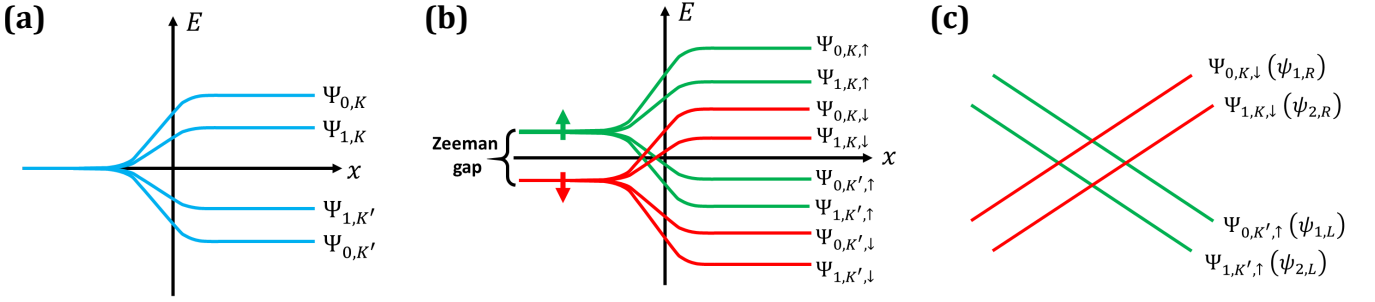}\\
  \caption{The boundary dispersion of the zero modes with spin degeneracy is shown in (a). When Zeeman effect is included, spin degenerate Landau levels split and the dispersion is shown in (b). In (c), we map our Landau level notation to the notation of effective theory used in the main text.}
  \label{Fig:Landau level}
\end{figure}

When Zeeman effect and spin degree of freedom are further taken into account, spin degeneracy is destroyed to form a bulk Zeeman gap. At the boundary, the in-gap edge modes form two pairs of helical edge mode as shown in Fig. \ref{Fig:Landau level} (b), whose gapless nature is protected by spin conservation symmetry and a spin Chern number of $2$. The mapping between Landau level edge states and the notation of chiral fermion used in the main text is shown in Fig. \ref{Fig:Landau level} (c).

\section{From Coulomb interaction to Luttinger parameter}
In the Landau gauge $A=(0,Bx,0)$, Landau level index $n$ and momentum $p$ along $y$ direction are both good quantum numbers. The real space electron operator $\Psi_n^{\dagger}(r)$ can be constructed
\bea
\Psi_n^{\dagger}(r)=\frac{1}{L}\sum_p e^{ipy}\phi_n(x-x_p)c_{n,p}^{\dagger}
\eea
where $c_{n,p}^{\dagger}$ creates an electron in the $n$th Landau level with momentum $p$. Here $\phi_n(x-x_p)\sim H_n(x-x_p)e^{-\frac{(x-x_p)^2}{2l_B^2}}$ up to a normalization factor and $H_n(x-x_p)$ is the $n$th Hermite polynomial. $L$ is the system size along $y$ direction and $x_p=pl_B^2$ is known as the position of guiding center. Notice that we are interested in the edge state physics which is localized on the boundary with a length scale of $l_B$. Therefore only LL states with a guiding center $x_p=0$ should be considered and this is equivalent to the long-wavelength limit where $p\sim \frac{1}{L}\sim 0$. As a result, the position $x$ and momentum in $y$ direction $p$ are decoupled in the LL wavefunction. The real space density operator is
\bea
\rho_n(r)=\Psi^{\dagger}_n(r)\Psi_n(r)=\frac{1}{L}|\phi_n(x)|^2\sum_q \rho_n(q)e^{iqy}
\eea
The standard Coulomb interaction is now written as,
\bea
&&\int d^2 r\int d^2r'\rho(r)V(r-r')\rho(r') \nonumber \\
&=&\frac{1}{L^2}\int d^2 r\int d^2r'|\phi_n(x)|^2|\phi_{n'}(x')|^2V(r-r')\sum_{q,q'}e^{iqy}e^{iq'y'}\rho(q)\rho(q') \nonumber \\
&=&\frac{1}{L^2}\int dx dx' d\tilde{y}|\phi_n(x)|^2|\phi_{n'}(x')|^2V(x-x',\tilde{y})\sum_{q,q'}[\int dye^{i(q+q')y}]\rho(q)\rho(q')e^{iq'\tilde{y}} \nonumber \\
&=&\frac{1}{L}\int_{-3l_B}^{3l_B} dx \int_{-3l_B}^{3l_B} dx' \int_{-\xi}^{\xi} d\tilde{y}|\phi_n(x)|^2|\phi_{n'}(x')|^2V(x-x',\tilde{y})\sum_{q}\rho(q)\rho(-q)e^{-iq\tilde{y}}
\eea
Here we have set the long distance cut-off of $x$ to be $3l_B$, which is a very good approximation as $\int_{-3l_B}^{3l_B}|\phi_{0,1}(x)|^2 dx\approx 0.999...$ The Coulomb potential takes the standard form $V(r-r')=\frac{e^2}{4\pi\epsilon\epsilon_0}\frac{1}{|r-r'|}$. In the long-wavelength limit, $e^{-iq\tilde{y}}\approx e^{-i\frac{\tilde{y}}{L}}\approx 1$. This is because $|\tilde{y}|$ ranges from atom spacing $a$ to Coulomb screening length $\xi$, and $\frac{\tilde{y}}{L}\ll1$. Therefore,
\bea
&&\int d^2 r\int d^2r'\rho_n(r)V(r-r')\rho_{n'}(r') \nonumber \\
&=&\frac{1}{L}\sum_{q}\rho_n(q)\rho_{n'}(-q)\int dx dx' d\tilde{y}|\phi_n(x)|^2|\phi_{n'}(x')|^2V(x-x',\tilde{y}) \nonumber \\
&=&g_{n,n'}\int dy \rho_n(y)\rho_{n'}(y).
\eea
We find that
\bea
g_{n,n'}=\int_{-3l_B}^{3l_B} dx \int_{-3l_B}^{3l_B}dx' \int_{-\xi}^{\xi}d\tilde{y}|\phi_n(x)|^2|\phi_{n'}(x')|^2V(x-x',\tilde{y})
\label{Eq:Luttinger-Coulomb relation}
\eea
where the short distance cut-off of Coulomb interaction is $a$. Now we are ready to evaluate the Luttinger parameter
\bea
K_+&=&\sqrt{\frac{v_f+\frac{1}{2\pi}(2g_5+g_3+g_4)}{v_f+\frac{1}{2\pi}(2g_5-g_3-g_4)}} \nonumber \\
K_-&=&\sqrt{\frac{v_f+\frac{1}{2\pi}(2g_5+g_3-g_4)}{v_f+\frac{1}{2\pi}(2g_5-g_3+g_4)}}
\eea
where we have considered the following density-density interaction
\bea
V_{density}=g_3\sum_{l=1,2}\rho_{l,L}\rho_{l,R}+g_4\sum_{l\neq l'\in\{1,2\}}\rho_{l,L}\rho_{l',R}+g_5 \sum_{l=1,2}(\rho_{l,L}^2+\rho_{l,R}^2)
\eea
With Eq. \ref{Eq:Luttinger-Coulomb relation}, we find that
\bea
g_3&=&g_{0,0}+g_{1,1} \nonumber \\
g_4&=&g_{0,1}+g_{1,0} \nonumber \\
g_5&=&2g_3
\eea
Assuming the screening length to be $\xi=1000a$, the out-of-plane magnetic field to be $2$ Tesla and a substrate with a dielectric constant $\epsilon=5$, we numerically calculate the value of Luttinger parameters, and find that
\bea
K_+=1.43,\ K_-=1.02.
\label{Eq:Luttinger parameter}
\eea

\section{An analytical estimation of Luttinger parameter}
Starting from Eq. \ref{Eq:Luttinger-Coulomb relation}, we can separate the integral along $y$ direction into two parts $g_{n,n'}=g_{n,n'}^>+g_{n,n'}^<$, where
\bea
g_{n,n'}^>&=&\int_{-3l_B}^{3l_B} dx \int_{-3l_B}^{3l_B}dx' \int_{3l_B\leq|\tilde{y}|\leq\xi}d\tilde{y}|\phi_n(x)|^2|\phi_{n'}(x')|^2V(x-x',\tilde{y}) \nonumber \\
g_{n,n'}^<&=&\int_{-3l_B}^{3l_B} dx \int_{-3l_B}^{3l_B}dx' \int_{a\leq|\tilde{y}|\leq 3l_B}d\tilde{y}|\phi_n(x)|^2|\phi_{n'}(x')|^2V(x-x',\tilde{y})
\eea
For $g_{n,n'}^>$, we take $\tilde{y}>3l_B\gg x-x'$ and find that
\bea
V(x-x',\tilde{y})&=&\frac{e^2}{4\pi\epsilon\epsilon_0}\frac{1}{\sqrt{(x-x')^2+\tilde{y}^2}} \nonumber \\
&\approx&\frac{e^2}{4\pi\epsilon\epsilon_0}\frac{1}{|\tilde{y}|}.
\eea
Therefore, the Coulomb interaction reduces to its one dimensional version under this limit, which gives rise to the following nice expression,
\bea
g_{n,n'}^>&\approx&\frac{e^2}{4\pi\epsilon\epsilon_0}\int_{-3l_B}^{3l_B} dx|\phi_n(x)|^2 \int_{-3l_B}^{3l_B}dx' |\phi_{n'}(x')|^2\int_{3l_B\leq|\tilde{y}|\leq\xi}d\tilde{y}\frac{1}{|\tilde{y}|} \nonumber \\
&=& \frac{e^2}{2\pi\epsilon\epsilon_0}\ln \frac{\xi}{3l_B},
\label{Eq:Analytical Luttinger parameter}
\eea
which reproduces the result demonstrated in the previous works \cite{kane1992}. It is now clear that $g_{n,n'}^>$ determines the one dimensional Coulomb interaction effect along the edge with $V(x-x',\tilde{y})=V(\tilde{y})$. If we first ignore $g_{n,n'}^<$, Eq. \ref{Eq:Analytical Luttinger parameter} offers us a rough while quick approach to estimate Luttinger parameters in this system. Notice that $g_{n,n'}^>$ in Eq. \ref{Eq:Analytical Luttinger parameter} is independent of the Landau level index $n$, so that $g_3=g_4$ and $K_-$ is not renormalized by the interactions. We find that
\bea
K_+(\text{analytical})=1.2,\ K_-(\text{analytical})=1.
\eea
The difference between this analytical result and numerical estimation in Eq. \ref{Eq:Luttinger parameter} reflects the two dimensional Coulomb corrections ($g_{n,n'}^<$), which originates from the finite penetration length of edge states.

\section{Correlation function of a (1+1) dimensional massive boson model}
In this section, we consider a (1+1) dimensional boson model with a large cosine potential of $\phi$, and show that when $\phi$ is pinned to the discrete minima, the correlation function of the dual field $\theta$ is diverging. Let us consider the following action,
\bea
S=\int d\tau dx \{i\partial_x \theta \partial_{\tau} \phi - \frac{v}{2}[K(\partial_x \phi)^2+\frac{1}{K}(\partial_x \theta)^2]-2g\cos \alpha \phi\}
\eea
In the large $g$ limit, we expand $\cos \alpha \phi$ to the second order and obtain
\bea
S=\int d\tau dx \{i\partial_x \theta \partial_{\tau} \phi - \frac{v}{2}[K(\partial_x \phi)^2+\frac{1}{K}(\partial_x \theta)^2]+\frac{m}{2} \phi^2\}
\eea
With Fourier transformation, we arrive at,
\bea
S=\frac{1}{2}\sum_q \Psi^{\dagger} G^{-1} \Psi
\eea
where
\bea
G^{-1}=
\frac{1}{\beta\Omega}
\begin{pmatrix}
vk^2K & ik\omega_n \\
ik\omega_n & \frac{vk^2}{K}+m \\
\end{pmatrix}
\eea
and $\Psi=(\theta_k,\phi_k)^T$. Then
\bea
G=\beta\Omega
\begin{pmatrix}
\frac{Km+vk^2}{Kk^2(vKm+v^2k^2+\omega_n^2)} & \frac{-i\omega_n}{k(vKm+v^2k^2+\omega_n^2)} \\
\frac{-i\omega_n}{k(vKm+v^2k^2+\omega_n^2)} & \frac{Kv}{vKm+v^2k^2+\omega_n^2}
\end{pmatrix}
\eea
The matrix $G$ tells us the correlation function in momentum space. For a time-ordered correlation function,
\bea
\langle {\cal T}[\phi(r)-\phi(0)]^2\rangle = \langle {\cal T}[\phi^2(r)+\phi^2(0)-\phi(r)\phi(0)-\phi(0)\phi(r)]\rangle
\eea
For $r=(x,\tau)$ and $q=(k,\omega_n)$, if $\tau>0$,
\bea
\langle {\cal T}[\phi(r)-\phi(0)]^2\rangle &=& \langle \phi^2(r)+\phi^2(0)-2\phi(r)\phi(0)\rangle \nonumber \\
&=&\frac{1}{(\beta\Omega)^2}\sum_q \langle \phi(q)\phi(-q)\rangle (2-2e^{iqr}) \nonumber \\
&=&\frac{2}{(\beta\Omega)^2}\sum_q \langle \phi(q)\phi(-q)\rangle (1-e^{ikx-i\omega_n|\tau|})
\eea
For $\tau<0$, $\tau=-|\tau|$, and
\bea
\langle {\cal T}[\phi(r)-\phi(0)]^2\rangle &=& \langle \phi^2(r)+\phi^2(0)-2\phi(0)\phi(r)\rangle \nonumber \\
&=&\frac{1}{(\beta\Omega)^2}\sum_q \langle \phi(q)\phi(-q)\rangle (2-2e^{-iqr}) \nonumber \\
&=&\frac{2}{(\beta\Omega)^2}\sum_q \langle \phi(q)\phi(-q)\rangle (1-e^{-ikx-i\omega_n|\tau|})
\eea
So generally, we find that
\bea
\langle {\cal T}[\phi(r)-\phi(0)]^2\rangle=\frac{2}{(\beta\Omega)^2}\sum_q \langle \phi(q)\phi(-q)\rangle (1-e^{isgn(\tau)kx-i\omega_n|\tau|})
\eea
In the zero temperature limit, the Matsubara frequency becomes continuous and can thus be integrated out,
\bea
\langle {\cal T}[\phi(r)-\phi(0)]^2\rangle&=&\frac{1}{2\pi^2}\int dk \int d\omega \frac{Kv}{m_0+v^2k^2+\omega^2} (1-e^{isgn(\tau)kx-i\omega|\tau|}) \nonumber \\
&=&\frac{vK}{2\pi^2}\int dk [\frac{\pi}{\sqrt{v^2k^2+m_0}}-\int d\omega \frac{e^{isgn(\tau)kx}}{\omega^2+(\sqrt{v^2k^2+m_0})^2}e^{-i\omega|\tau|} ] \nonumber \\
&=&\frac{vK}{2\pi}\int dk\frac{1}{\sqrt{v^2k^2+m_0}}(1-e^{isgn(\tau)kx}e^{-\sqrt{v^2k^2+m_0}|\tau|})
\eea
where we have defined $m_0=vKm$. Therefore, we can get rid of the sign function and arrive at,
\bea
\langle {\cal T}[\phi(r)-\phi(0)]^2\rangle=\frac{vK}{\pi}\int_0^{\infty} dk\frac{1}{\sqrt{v^2k^2+m_0}}(1-\cos (kx) e^{-\sqrt{v^2k^2+m_0}|\tau|})
\label{Eq:Correlation phi}
\eea
Similarly, one can show that the correlation function for the dual field is given by,
\bea
\langle {\cal T}[\theta(r)-\theta(0)]^2\rangle=\frac{1}{\pi vK}\int_0^{\infty} dk\frac{\sqrt{v^2k^2+m_0}}{k^2}(1-\cos (kx) e^{-\sqrt{v^2k^2+m_0}|\tau|})
\label{Eq:Correlation theta}
\eea
When $m_0$ is set to zero, we recover the well-known result of logarithmic correlation of Luttinger liquid. When $m_0\rightarrow \infty$, $\phi$ is pinned to constant value, and thus $\langle {\cal T}[\phi(r)-\phi(0)]^2\rangle\rightarrow 0$. On the other hand, $\phi$ and $\theta$ are conjugate to each other. So the quantum fluctuation of $\theta$ should lead to the divergence of the correlation function as $m_0\rightarrow \infty$ limit. To see this, let us consider a momentum cut-off $\Lambda$ $k\leq \Lambda\ll m_0$ and expand the correlation function of $\theta$ to the second order,
\bea
\langle {\cal T}[\theta(r)-\theta(0)]^2\rangle &>& \frac{1}{\pi vK}\int_0^{\Lambda} dk\frac{\sqrt{v^2k^2+m_0}}{k^2}(1-\cos (kx) e^{-\sqrt{v^2k^2+m_0}|\tau|}) \nonumber \\
&>& \frac{\sqrt{m_0}}{\pi vK}\int_0^{\Lambda} dk\frac{1}{k^2}(1-\cos (kx) e^{-\sqrt{m_0}|\tau|}) \nonumber \\
&\sim& \frac{\sqrt{m_0}}{\pi vK}\int_0^{\Lambda} dk\frac{1}{k^2}
\eea
Therefore, $\langle {\cal T}[\theta(r)-\theta(0)]^2\rangle$ is diverging in two different ways: As $k\rightarrow 0$, it has an infrared divergence. Meanwhile, its value scales with the mass $\sqrt{m_0}$.

\section{Phase diagram in a two-channel QSH system}
In this section, we discuss the phase diagram of bilayer graphene QSH in a QPC. First of all, the single particle tunneling is
\bea
T_{l,l'}=t_{l,l'}\psi^{\dagger}_{t,l,L}\psi_{b,l',R}+h.c.\sim t_{l,l'}\cos \sqrt{\pi}(\phi_{t,l}+\theta_{t,l}+\phi_{b,l'}-\theta_{b,l'})
\eea
The scaling dimension of this term is
\bea
\Delta(t_{l,l'})=\frac{1}{4}(K_l+\frac{1}{K_l}+K_{l'}+\frac{1}{K_{l'}})\geq 1
\eea
Therefore, single particle tunneling is always marginal ($K_l=K_{l'}=1$) or irrelevant under RG. For spin-1 tunneling, its bosonized formula is
\bea
V^{\sigma}_{l_1,l_2,l_3,l_4}&=&v^{\sigma}_{l_1,l_2,l_3,l_4}\psi^{\dagger}_{b,L,l_1}\psi_{t,R,l_2}\psi^{\dagger}_{t,L,l_3}\psi_{b,R,l_4}+h.c. \nonumber \\
&=&v^{\sigma}_{l_1,l_2,l_3,l_4} \cos \sqrt{\pi} [(\phi_{b,l_1}+\phi_{b,l_4})+(\phi_{t,l_2}+\phi_{t,l_3})+(\theta_{b,l_1}-\theta_{b,l_4})+(-\theta_{t,l_2}+\theta_{t,l_3})]
\eea
Next, we hope to exhaust the choices of channel indices, and check the scaling dimensions of different spin-1 tunneling process. Typical examples of different channel choices are shown below. For $l_1=l_2=l_3=l_4=l_0$ with $l_0=I/II$, and
\bea
V^{\sigma}_1(l_0)&=&v^{\sigma}_1\cos 2\sqrt{\pi} (\phi_{+,+}-(-1)^{l_0}\phi_{+,-}) \nonumber \\
\Delta(v^{\sigma}_1)&=&\frac{1}{K_+}+\frac{1}{K_-}
\eea
For $l_1=l_4\neq l_2=l_3=l_0$,
\bea
V^{\sigma}_2(l_0)&=&v^{\sigma}_2\cos 2\sqrt{\pi} (\phi_{+,+}-(-1)^{l_0}\phi_{-,-}) \nonumber \\
\Delta(v^{\sigma}_2)&=&\frac{1}{K_+}+\frac{1}{K_-}
\eea
For $l_1=l_2\neq l_3=l_4=l_0$,
\bea
V^{\sigma}_3(l_0)&=&v^{\sigma}_3\cos 2\sqrt{\pi} (\phi_{+,+}-(-1)^{l_0}\theta_{-,-}) \nonumber \\
\Delta(v^{\sigma}_3)&=&\frac{1}{K_+}+K_-
\eea
The tunneling term with $l_1=l_3\neq l_2=l_4=l_0$ shares the same scaling dimension with $V^{\sigma}_3(l_0)$.
For $l_1=l_2=l_3\neq l_4=l_0$,
\bea
V^{\sigma}_4(l_0)&=&v^{\sigma}_4\cos \sqrt{\pi} [2\phi_{+,+}+(-1)^{l_0}(\phi_{+,-}+\phi_{-,-})+(-1)^{l_0}(\theta_{+,-}-\theta_{-,-})] \nonumber \\
\Delta(v^{\sigma}_4)&=&\frac{1}{K_+}+\frac{1}{2}(K_-+\frac{1}{K_-})\geq 1+\frac{1}{K_+}>1
\eea
One can generally show that a spin-1 tunneling process that has three channel indices equal to each other shares the same scaling dimension of $V^{\sigma}_4(l_0)$. These terms are irrelevant under RG as $\Delta(v^{\sigma}_4)>1$ despite the value of $K_{\pm}$. Therefore, spin-1 tunneling is relevant when either $\frac{1}{K_+}+\frac{1}{K_-}<1$ or $\frac{1}{K_+}+K_-<1$ is satisfied.

For charge-2e tunneling, we have
\bea
V^{\rho}_{l_1,l_2,l_3,l_4}&=&v^{\rho}_{l_1,l_2,l_3,l_4}\psi^{\dagger}_{b,L,l_1}\psi_{t,R,l_2}\psi^{\dagger}_{b,R,l_4}\psi_{t,L,l_3}+h.c. \nonumber \\
&=&v^{\rho}_{l_1,l_2,l_3,l_4} \cos \sqrt{\pi} [(\phi_{b,l_1}-\phi_{b,l_4})+(\phi_{t,l_2}-\phi_{t,l_3})+(\theta_{b,l_1}+\theta_{b,l_4})-(\theta_{t,l_2}+\theta_{t,l_3})]
\eea
Similarly, for $l_1=l_2=l_3=l_4=l_0$ with $l_0=I/II$, and
\bea
V^{\rho}_1(l_0)&=&v^{\rho}_1\cos 2\sqrt{\pi} (\theta_{-,+}-(-1)^{l_0}\theta_{-,-}) \nonumber \\
\Delta(v^{\sigma}_1)&=&K_+ +K_-
\eea
For $l_1=l_4\neq l_2=l_3=l_0$,
\bea
V^{\rho}_2(l_0)&=&v^{\rho}_2\cos 2\sqrt{\pi} (\theta_{-,+}+(-1)^{l_0}\theta_{+,-}) \nonumber \\
\Delta(v^{\sigma}_2)&=&K_+ +K_-
\eea
For $l_1=l_2\neq l_3=l_4=l_0$,
\bea
V^{\rho}_3(l_0)&=&v^{\rho}_3\cos 2\sqrt{\pi} (\phi_{+,-}-(-1)^{l_0}\theta_{-,+}) \nonumber \\
\Delta(v^{\rho}_3)&=&K_+ +\frac{1}{K_-}
\eea
The tunneling term with $l_1=l_3\neq l_2=l_4=l_0$ shares the same scaling dimension with $V^{\rho}_3(l_0)$.
For $l_1=l_2=l_3\neq l_4=l_0$,
\bea
V^{\rho}_4(l_0)&=&v^{\rho}_4\cos \sqrt{\pi} [2\theta_{-,+}+(-1)^{l_0}(\phi_{+,-}-\phi_{-,-})-(-1)^{l_0}(\theta_{+,-}+\theta_{-,-})] \nonumber \\
\Delta(v^{\rho}_4)&=&K_+ +\frac{1}{2}(K_-+\frac{1}{K_-})\geq K_+ + 1>1
\eea
Similarly, all the terms with three equal indices are irrelevant under RG as $\Delta(v^{\rho}_4)>1$. To conclude, charge-2e tunneling processes are relevant when either $K_+ +\frac{1}{K_-}<1$ or $K_++K_-<1$ is satisfied. There also exists a region where neither IC nor CI phase are favored, where CC/II phase is stable. This concludes the phase diagram of QSH state, which is summarized in Fig. 3 (a) in the main text.


\section{Instanton tunneling and the effective charge}
A direct evidence of BSPT state is to extract bosonic $2e$-charge from shot noise spectrum, as proposed in our earlier work \cite{bi2016}. The QPC setup offers us an ideal platform to perform noise measurement. In the following two sections, we give a detailed analysis about the origin of bosonic effective charge and confirm its existence in the noise spectrum. We first start with an analysis of instanton physics in the IC fixed point, and demonstrate that how $e$-instanton charge in QSH state is modified to $2e$-instanton charge when BSPT condition is applied. We then move to calculate the shot noise spectrum of instanton current using non-equilibrium Keldysh formulism, and demonstrate that $2e$ bosonic charge can be extracted from this noise measurement.

In the IC limit, spin-1 tunneling $V^{\sigma}$ is large and relevant. $\phi_{\sigma}$ is pinned to the minima of cosine potential and lose its dynamics. However, quantum tunneling between neighboring minima of $V^{\sigma}$ generally exists and is identified as instanton tunneling process. For QSH state, a typical example of $V^{\sigma}$ term is 
\bea
V^{\sigma}_1(l_0)&=&v^{\sigma}_1(l_0)\psi^{\dagger}_{b,L,l_0}\psi_{t,R,l_0}\psi^{\dagger}_{t,L,l_0}\psi_{b,R,l_0}+h.c. \nonumber \\
&\sim& v^{\sigma}_1(l_0)\cos 2\sqrt{\pi}(\phi_{+,+}-(-1)^{l_0}\phi_{+,-}),
\eea
where $l_0=I,II$ . This term is relevant when $\frac{1}{K_+}+\frac{1}{K_-}<1$, where $\phi_{+,+}=\phi_{\rho}=\frac{(n+m+1)\pi}{2\sqrt{\pi}}$ and $\phi_{+,-}=\frac{(n-m)\pi}{2\sqrt{\pi}}$ are pinned to certain discrete values with $n,m\in \mathbb{Z}$. Instanton tunneling happens when either $n$ or $m$ changes, while it is worth noticing that only a change of $\phi_{\rho}$ is related to electric charge process. In particular, the minimal instanton tunneling of $\phi_{\rho}$ is $\Delta\phi_{\rho}=\frac{\sqrt{\pi}}{2}$, which corresponds to 
\bea
\Delta n+\Delta m=1.
\label{Eq:QSH IT}
\eea

\begin{figure}
  \centering
  \includegraphics[width=3.5in]{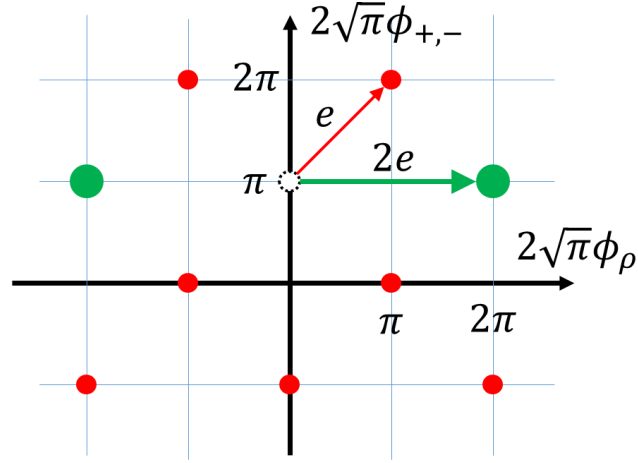}\\
  \caption{Minimal instanton process of BSPT (QSH) IC phase is demonstrated by a green (red) arrow, characterizing $2e$ ($e$) charge signal in the noise spectrum.}
  \label{Fig:Instanton}
\end{figure}

As shown in Fig. \ref{Fig:Instanton}, instanton tunneling of QSH IC phase corresponds to the hopping from the open circle to the colored circles in the configuration space of $\phi_{\rho}$ and $\phi_{+,-}$, and a minimal instanton tunneling is depicted by the red arrow. Physically, the electric charge of the minimal instanton tunneling is
\bea
\Delta Q=e\int dt I_{\rho}=-\frac{2e}{\sqrt{\pi}}\Delta\phi_{\rho}=-e.
\eea
Because charge current is $I_{\rho}=I_1+I_4-I_2-I_3$, the minimal instanton tunneling pumps $e$-charge from left (lead 1 and 4) to right (lead 2 and 3) across the QPC. One can easily show that the minimal instanton tunnelings of other spin-1 tunneling terms give exactly the same charge-e pumping process. Although here we focus on the IC phase of QSH state to compare with that of BSPT state, this $e$-charge feature should be generally shared by other phases of QSH state, as a result of its fermionic nature.

For the same $V^{\sigma}$ in the BSPT limit, however, $\phi_{t/b,-}$ is pinned to be a constant, and will not participate the instanton physics. This leads to an additional constraint 
\bea
\Delta n=\Delta m.
\label{Eq:BSPT IT}
\eea
Consequently, the charge-$e$ instanton process characterized by Eq. \ref{Eq:QSH IT} is forbidden, since $\Delta n, \Delta m\in \mathbb{Z}$. The new minimal instanton process, which is consistent with Eq. \ref{Eq:BSPT IT}, is depicted by the green arrow in Fig. \ref{Fig:Instanton} (c). By calculating the tunneling charge, it is easy to show that this process corresponds to a charge-$2e$ pumping from left to right. Therefore, the difference between the $2e$-instanton of BSPT and the $e$-instanton of QSH originates from the BSPT condition, which reflects the bosonic nature of BSPT physics.

\section{Noise spectrum}
Experimentally, we expect that a shot noise measurement is able to identify the instanton charge and thus distinguishes the BSPT from a QSH state in bilayer graphene. In the this section, we will calculate the relation between the non-equilibrium current and the non-equilibrium shot noise spectrum, and extract the effective charge from this relation \cite{martin2005,maciejko2009}. To start with, we first derive the expression of the equilibrium current of instanton process in the following action
\bea
S=\frac{K}{\beta}\sum_{n}|\omega_n||\phi(\omega_n)|^2+g\int d\tau \cos [C\sqrt{\pi}\phi (\tau)],
\eea
where the value of $C$ depends on the details of interaction. Here we have integrated out the field at $x\neq 0$ and arrive at the $0+1$ dimensional action at $x=0$. When $\frac{C^2}{4K}<1$, the cosine potential is relevant and flows to strong coupling limit under RG. The $\phi$ field will be pinned to $\frac{(2n+1)}{C}\sqrt{\pi}$ ($n\in\mathbb{Z}$) to minimize the free energy of the system. Then a single instanton process that happens at $\tau=\tau_i$ can be described as
\bea
\phi(\tau)=\phi_0+q_i\frac{2\sqrt{\pi}}{C}\Theta(\tau-\tau_i).
\eea
Here $\phi_0$ is the initial field configuration at $\tau=0$ and $q_i=\pm 1$ is the charge of the instanton.

Physically, the charge of instanton must satisfy the charge neutrality condition as a result of the periodic boundary condition $\phi(\tau=0)=\phi(\tau=\beta)$. Therefore, a multiple instanton configuration should be considered, where
\bea
\phi=\phi_0+\frac{2\sqrt{\pi}}{C}\sum_i q_i\Theta(\tau-\tau_i)
\eea
and $\sum_i q_i=0$.
Then the equilibrium current operator for the instanton process is
\bea
\langle j(\tau)_{eq}\rangle =\frac{ieD}{\sqrt{\pi}}\langle \partial_{\tau} \phi\rangle_S=i\frac{2eD}{C}\langle \sum_i q_i\delta(\tau-\tau_i) \rangle_S
\eea
where we have used
\bea
\frac{d\phi}{d\tau}=\frac{2\sqrt{\pi}}{C}\sum_i q_i\delta(\tau-\tau_i)
\eea
and $D$ is another adjustable parameter, which depends on the model details. On the other hand, the dual action of $S$ which describes the instanton process in the strong coupling limit is
\bea
S_{dual}=\frac{1}{\beta K}\sum_{\omega_n}|\omega_n||\tilde{\phi}|^2+2\tilde{g}\int d\tau\cos \frac{4}{C}\sqrt{\pi}\tilde{\phi}
\eea
$\tilde{\phi}$ here is the dual field of $\phi$. Further, we can define an action $S[\phi,a]_{dual}$ with an auxiliary gauge field $a(\tau)$:
\bea
S[\phi,a]_{dual}=\frac{1}{\beta K}\sum_{\omega_n}|\omega_n||\tilde{\phi}|^2+2\tilde{g}\int d\tau\cos [\frac{4}{C}\sqrt{\pi}\tilde{\phi}+\frac{2eD}{C}a(\tau)]
\eea
In the Coulomb gas formulism, the partition function of $S[\phi,a]_{dual}$ is now
\bea
\frac{Z^{dual}}{Z^{dual}_0}=1+\sum_{p}^{\infty}\frac{\tilde{g}^{2p}}{2p!}\prod_{i=1}^{2p}\sum_{q_i=\pm1} \int_0^{\beta} d\tau_i exp[\frac{8 K}{C^2}\ln \frac{|\tau_i-\tau_j|}{\tau_c}+i\sum_iq_i\frac{2eD}{C}a(\tau_i)]
\eea
where $Z^{dual}_0$ is short for $Z^{dual}(a=0)$. The current can be given by
\bea
\frac{1}{Z^{dual}_0}\frac{\delta Z^{dual}}{\delta a}|_{a\rightarrow 0}=i\frac{2eD}{C}\langle\sum_iq_i\delta(\tau-\tau_i)\rangle=\langle j(\tau) \rangle_{eq}
\eea
Therefore, the equilibrium current is
\bea
\langle j(\tau) \rangle_{eq}= \frac{1}{Z^{dual}_0}\frac{\delta Z^{dual}}{\delta a}|_{a\rightarrow 0} = \frac{4e\tilde{g}D}{C}\sin [\frac{4}{C}\sqrt{\pi}\tilde{\phi}]
\eea
If a finite bias $V$ is applied, we only need to replace $a(\tau)$ with the electromagnetic vector potential $A(\tau)$ in the above formula, where $A=-Vt$. In this case, the current operator is
\bea
\langle j(\tau) \rangle_{eq}=\frac{4e\tilde{g}D}{C}\sin [\frac{4}{C}\sqrt{\pi}\tilde{\phi}+\frac{2eD}{C}A(\tau)]
\eea

Let us go back to real time $t$. The non-equilibrium current is obtained using the Keldysh technique. Define the Keldysh contour as $K$ and an index $\eta=\pm$ characterizing the forward ($+$) and backward ($-$) branch. The non-equilibrium current is
\bea
\langle I(t) \rangle=\frac{1}{2}\sum_{\eta}\langle {\cal T}_K j(t^{\eta})_{eq} e^{-i\int_K dt_1 H_1(t_1)} \rangle
\eea
where $H_1=-2\tilde{g}\int dt\cos [\frac{4}{C}\sqrt{\pi}\tilde{\phi}+\frac{2eD}{C}A(t)]$ is the perturbation term in the Hamiltonian formulism. Notice that $\tilde{g}$ is a small parameter in the dual theory, and we expand the current to the leading order,
\bea
\langle I(t) \rangle &=& \frac{4e\tilde{g}^2D}{C}\sum_{\eta}\langle {\cal T}_K \sin [\frac{4}{C}\sqrt{\pi}\tilde{\phi}+\frac{2eD}{C}A(t)] (1+i\int_K dt_1 \cos [\frac{4}{C}\sqrt{\pi}\tilde{\phi}+\frac{2eD}{C}A(t_1)]) \rangle+{\cal O}(\tilde{g}^2) \nonumber \\
&=& i\frac{4e\tilde{g}^2D}{C}\int dt_1 \sum_{\eta,\eta_1}\eta_1 \langle {\cal T}_K \sin [\frac{4}{C}\sqrt{\pi}\tilde{\phi}(t)+\frac{2eD}{C}A(t)] \cos [\frac{4}{C}\sqrt{\pi}\tilde{\phi}(t_1)+\frac{2eD}{C}A(t_1)] \rangle+{\cal O}(\tilde{g}^2)
\eea
By making use of the correlation function properties of vertex operators, it is straightforward to show that
\bea
&&\langle {\cal T}_K \sin [\frac{4}{C}\sqrt{\pi}\tilde{\phi}(t)+\frac{2eD}{C}A(t)] \cos [\frac{4}{C}\sqrt{\pi}\tilde{\phi}(t_1)+\frac{2eD}{C}A(t_1)] \rangle \nonumber \\
&=& \frac{1}{2}\sin [\frac{2eD}{C}(A(t)-A(t_1))]\langle {\cal T}_K e^{i\frac{4}{C}\sqrt{\pi}\tilde{\phi}(t)}e^{-i \frac{4}{C}\sqrt{\pi}\tilde{\phi}(t_1)} \rangle \nonumber \\
&=& \frac{1}{2}\sin [\frac{2eDV}{C}(t-t_1)] e^{-\frac{8\pi}{C^2}\langle {\cal T}_K [\tilde{\phi}(t)-\tilde{\phi}(t_1)]^2\rangle}
\eea
So the non-equilibrium current is
\bea
\langle I(t) \rangle =i\frac{4e\tilde{g}^2D}{C}\int_{-\infty}^{\infty} dt_1 \sum_{\eta,\eta_1}\eta_1 \frac{1}{2}\sin [\frac{2eDV}{C}(t-t_1)] e^{-\frac{8\pi}{C^2}\langle {\cal T}_K [\tilde{\phi}(t)-\tilde{\phi}(t_1)]^2\rangle}
\eea
In the Keldysh formulism, we have four different correlation functions $G_{\pm,\pm}$. Notice that in the above formula, if a Green function $G_{\eta,\eta'}$ is even in time, it gives zero contribution to the non-equilibrium current. So we only need to consider the following green function
\bea
D_{+,-}(t)&=&\frac{K}{2\pi}\ln \frac{\pi\tau_c/\beta}{\sin [\pi(-it)/\beta]} \nonumber \\
D_{-,+}(t)&=&\frac{K}{2\pi}\ln \frac{\pi\tau_c/\beta}{\sin [\pi(it)/\beta]}
\eea
Then
\bea
\langle I(t) \rangle &=&i\frac{2e\tilde{g}^2D}{C}\int_{-\infty}^{\infty} dt_1 \sum_{\eta,\eta_1}\eta_1 \sin [\frac{2eDV}{C}(t-t_1)] exp(\frac{16\pi}{C^2}\frac{K}{2\pi}\ln \frac{\pi\tau_c/\beta}{\sin [\pi(\eta_1 i)(t-t_1)/\beta]}) \nonumber \\
&=&i\frac{2e\tilde{g}^2D}{C}\int_{-\infty}^{\infty} dt \sum_{\eta}(-\eta) \sin [\frac{2eDV}{C}t] exp(\frac{8K}{C^2}\ln \frac{\pi\tau_c/\beta}{\sin [\pi(\eta it)/\beta]}) \nonumber \\
&=&i\frac{2e\tilde{g}^2D}{C}(\frac{\pi\tau_c}{\beta})^{\frac{8K}{C^2}}\int_{-\infty}^{\infty} dt \sum_{\eta}(-\eta) \sin [\frac{2eDV}{C}t] [\frac{1}{-\eta i\sinh [\pi t/\beta]}]^{\frac{8K}{C^2}}
\eea
This integral can be calculated exactly when $\frac{8K}{C^2}<1$, while a singularity occurs at $t=0$ when $\frac{8K}{C^2}\geq 1$.

The non-equilibrium noise spectrum at zero frequency $\tilde{S}(\omega=0)$ is defined as
\bea
\tilde{S}(\omega=0)=\int d(t-t')\tilde{S}(t-t')
\eea
While $\tilde{S}(t-t')$ is defined in the Keldysh formulism as
\bea
\tilde{S}(t-t')&=&\sum_{\eta}\langle {\cal T}_K \{I_{\eta}(t),I_{-\eta}(t')e^{-i\int_K dt_1 H_1(t_1)}\} \rangle \nonumber \\
&=&\sum_{\eta}\langle {\cal T}_K \{I_{\eta}(t),I_{-\eta}(t')\}\rangle + {\cal O}(\tilde{g}^2 )
\eea
Following the calculation of non-equilibrium current, we obtain
\bea
\tilde{S}(t)&=&\frac{1}{2}(\frac{4e\tilde{g}D}{C})^2\sum_{\eta}\cos [\frac{2eDV}{C}t] e^{\frac{16\pi}{C^2}D_{\eta,-\eta}(t)} \nonumber \\
&=&\frac{1}{2}(\frac{4e\tilde{g}D}{C})^2(\frac{\pi \tau_c}{\beta})^{\frac{8K}{C^2}}\sum_{\eta}\cos [\frac{2eDV}{C}t] (\frac{1}{-\eta i\sinh[\pi t/\beta]})^{\frac{8K}{C^2}} \nonumber \\
\eea
where we have defined $t-t'\rightarrow t$. Therefore, the non-equilibrium noise spectrum at zero frequency is
\bea
\tilde{S}(\omega=0)=\frac{1}{2}(\frac{4e\tilde{g}D}{C})^2(\frac{\pi \tau_c}{\beta})^{\frac{8K}{C^2}}\sum_{\eta}\int dt \cos [\frac{2eDV}{C}t] (\frac{1}{-\eta i\sinh[\pi t/\beta]})^{\frac{8K}{C^2}}
\eea
Therefore, we arrive at a similar diverging integral. Let us denote these two integrals as
\bea
F_1(a,b,c)&=&\sum_{\eta}(-\eta) \int_{-\infty}^{\infty} dt \sin at (\frac{1}{-i\eta\sinh bt})^c \nonumber \\
F_2(a,b,c)&=&\sum_{\eta} \int_{-\infty}^{\infty} dt \cos at (\frac{1}{-i\eta\sinh bt})^c
\eea
where $a=\frac{2eV}{C}, b=\frac{\pi}{\beta}, c=\frac{8K}{C^2}$. To evaluate these integrals, we perform the analytic continuation and define,
\bea
\tau=-t+\frac{i\eta\pi}{2b},\ t=-\tau+\frac{i\eta\pi}{2b}
\eea
This shift allows us to define a new integration contour in the complex plane. It can be checked that between the old contour and the new contour, there is no singularity. Therefore, according to Cauchy's theorem, the new integration equals to the old integration, while there is no singularity along the new contour at all. Then we find that
\bea
\sinh (b\tau-\frac{i\eta\pi}{2})&=&-i\eta\cosh b\tau
\eea
now becomes an even function in $\tau$. Therefore, after this transformation, any $\tau$-odd component of $\sin at$ and $\cos at$ must vanish under the integration. Then we arrive at
\bea
F_1(a,b,c)&=&2i\sinh \frac{\pi a}{2b}\int_{-\infty}^{\infty}d\tau \frac{\cos a\tau}{(-\cosh b\tau)^c} \nonumber \\
F_2(a,b,c)&=&-2\cosh \frac{\pi a}{2b}\int_{-\infty}^{\infty}d\tau \frac{\cos a\tau}{(-\cosh b\tau)^c}
\eea
This gives rise to an interesting relation between these two integrals that
\bea
F_2(a,b,c)=i\coth \frac{\pi a}{2b} F_1(a,b,c)
\eea
Without evaluating the integral explicitly, we arrive at the following relation between noise spectrum and current,
\bea
\tilde{S}(\omega=0)=2e^*\coth \frac{e^*V}{Ck_BT} \langle I \rangle
\eea
where the effective charge is
\bea
e^*=\frac{2De}{C}
\eea
For a quantum point contact in a BSPT system. In the IC phase, $D=C=2$ and we find that the effective charge is $e^*=2e$, which is consistent with the instanton analysis in the main text.

\end{document}